\documentclass[twocolumn,showpacs,preprintnumbers,prb]{revtex4}
\usepackage{dcolumn}
\usepackage{amsmath}
\usepackage{amssymb}
\usepackage{bm}
\usepackage{graphicx}
\usepackage{float}
\newcommand{\avg}[1]{\left< #1 \right>} 
\newcommand{\ket}[1]{\left| #1 \right>} 
\newcommand{\bra}[1]{\left< #1 \right|} 

\DeclareRobustCommand{\orderof}{\ensuremath{\mathcal{O}}}

\begin{document}

\title{Effect of Single Impurity on Free Fermion Entanglement Entropy
}

\author{Mohammad Pouranvari and Kun Yang}
\affiliation{National High Magnetic Field Laboratory and Department of Physics,
Florida State University, Tallahassee, Florida 32306, USA}
\author{Alexander Seidel}
\affiliation{Department of Physics, Washington University, St. Louis, MO 63160, USA}
\date{\today}

\begin{abstract}
The one-dimensional free Fermi gas is a prototype conformally invariant system, whose entanglement properties are well-understood. In this work, the effects of a single impurity on one dimensional free fermion entanglement entropy are studied both analytically and numerically. Such an impurity represents an exactly marginal perturbation to the bulk conformally invariant fixed point. We find that the impurity leads to sub-leading contributions to the entanglement entropy that scale inversely with the subsystem size. The origin of such contributions are identified.
\end{abstract}

\maketitle

\section{Introduction}

Entanglement is emerging as an important characteristic of many-particle systems. While not yet a directly accessible quantity experimentally, entanglement entropy (EE) and
spectrum have been studied theoretically and numerically in many different models. In general they are {\em not} simply related to correlation functions measured in linear
response, and thus provide alternative probes of the system that are particularly useful in determining the nature of phases and phase transitions.\cite{lecturereview}

Among the many systems whose entanglement properties have been studied, the best understood are perhaps one-dimensional (1D) conformally-invariant systems. In this case
it is not only understood that bipartite EE grows logarithmically with subsystem size with a coefficient depending on the central charge
only,\cite{cardyreview} but also  much is known about how a single impurity affects EE.\cite{affleckreview} The effect of the impurity depends sensitively
on whether the perturbation it induces is relevant or not in the renormalization group (RG) sense. When it is relevant, it effectively cuts the chain into two decoupled
pieces in the scaling limit, and typically produces a finite correction to entanglement entropy. On the other hand, when it is irrelevant, its effect goes to zero in a
power-law fashion in the scaling limit.\cite{erikson} What has been left out in previous studies is the case when the impurity is {\em exactly marginal}. This is the
case when a single potential scatterer is embedded in a 1D free Fermi gas,\cite{KaneFisher92} which is the subject of the present work. The non-interacting nature of the system allows
for detailed numerical studies in sufficiently large systems sizes that yields results in the scaling regime.

Our main findings are the following. A single impurity in a 1D free fermion lattice induces two subleading contributions to EE, both of which scales as $1/N_A$, where $N_A$ is the subsystem size. One of these contributions oscillates with the Fermi wave vector, while the other is non-oscillatory. The origin of both terms are identified by considering certain limiting cases, where analytic treatment is available. We note that a closely related recent work\cite{ossipov} did not find any corrections to EE for such impurities.

The rest of the paper is organized as follows. In section \ref{analytic} we show by analytic calculation that an inversion symmetric potential adds to the continuum free Fermi gas EE a term inversely proportional to the subsystem size. We verify this numerically in the succeeding sections for a discrete lattice system. In section \ref{method} we introduce the one dimensional single impurity Hamiltonian model and the numerical method of
calculating the EE. In section \ref{numeric}, detailed numerical calculations are presented; we first verify numerically asymptotic and subleading behavior of the free fermion EE up to order of $1/N_A^2$. Then, we show that the single impurity introduces a term on the order of $1/N_A$ to the EE. In the rest of this section, behavior of the subleading term is  examined  numerically in the regime of large impurity strength, small impurity strength, and disorder on the order $1$. Some concluding remarks are offered in section \ref{con}.

\section{Analytic Approach}\label{analytic}
Let us consider a 1D free Fermi gas with states $-k_F < k < k_F$ fully occupied, and the entanglement between a subsystem A defined by $-L < x < L$ with the rest of the infinite line (called B). All entanglement-related properties are encoded in the two overlap matrices:
\begin{eqnarray}
A^e_{kk'}=\int_0^{L}\cos(kx)\cos(k'x)dx,\\
A^o_{kk'}=\int_0^{L}\sin(kx)\sin(k'x)dx,
\end{eqnarray}
where we used the fact that there is no overlap between even and odd functions. Equivalently, but perhaps done less often, we can work with the analog of the above in subsystem B
\begin{eqnarray}
B^e_{kk'}=\int_L^{\infty}\cos(kx)\cos(k'x)dx,\\
B^o_{kk'}=\int_L^{\infty}\sin(kx)\sin(k'x)dx.
\end{eqnarray}
Now introduce a inversion symmetric impurity potential $V(x)=V(-x)$ with range $R$ such that $V(x>R)=0$. There are two special features of such potential that are very useful for later considerations: (i) Energy eigenstates will remain parity eigenstates, such that there is no mixing between even and odd parity states. (ii) As long as $L > R$, the only effect on the wavefunctions in subsystem B is a phase shift: $\cos(kx)\rightarrow\cos(kx-\delta^e_k)$, and $\sin(kx)\rightarrow\sin(kx-\delta^o_k)$. It is therefore advantageous to work with the B matrices as they only involve the phase shifts but not the complicated wavefunction in the region $|x| < R$ where it is not a plane wave. Furthermore, in the low-energy limit $k_FR \ll 1$, the $k$-dependence of the phase shifts can be parameterized by scattering lengths:
\begin{eqnarray}
\delta^{e,o}_k\approx k \mathfrak{a}^{e,o}.
\label{eq:PhaseShiftInTermsOfScatteringLength}
\end{eqnarray}
In the asymptotic limit where this behavior becomes exact, we find
\begin{eqnarray}
B^e_{kk'}=\int_L^{\infty}\cos(kx-k \mathfrak{a}^e)\cos(k'x-k' \mathfrak{a}^e)dx\nonumber\\
=\int_{L- \mathfrak{a}^e}^{\infty}\cos(kx)\cos(k'x)dx,\label{Bkk}
\end{eqnarray}
namely, the entanglement properties of the even sector is identical to those of the original impurity free case with $L'=L- \mathfrak{a}^e$. Similarly the entanglement properties of the odd sector is identical to those of the original impurity free case with $L'=L- \mathfrak{a}^o$. In particular, we expect the entanglement entropy for the impurity free case to behave as
\begin{eqnarray}
EE(k_F, L)={1\over 3}\ln(k_F L)+O(1);
\end{eqnarray}
the considerations above immediately suggest that the impurity will lead to a correction
\begin{eqnarray} \label{deltaEE}
\delta EE(k_F, L)=-{ \mathfrak{a}^e+ \mathfrak{a}^o\over 6L}+O(1/L^2).
\label{eq:ContinuumEEChange}
\end{eqnarray}

We  point out that equations such as \eqref{Bkk} receive a contribution proportional
to $\delta(k \pm k')$ coming from the oscillation at infinity and which is independent of $L$ and the
scattering length(s). On top of that, however, there is a finite contribution that is
in general non-vanishing for different $k$, $\pm k'$, owing to the fact that cutting off
the integration at $L$ spoils the orthogonality of the factors  in the integrand.
It is this finite piece that contains information about the dependence of the
EE on $L$ and the scattering lengths.
We also note that Eq. (\ref{eq:PhaseShiftInTermsOfScatteringLength}) is exact only for a hard wall potential that excludes the particle from $-\mathfrak{a} < x < \mathfrak{a}$, in which case Eq. (\ref{eq:ContinuumEEChange}) becomes exact. For generic potentials there are additional corrections to the  EE not captured by Eq. (\ref{eq:PhaseShiftInTermsOfScatteringLength}), as we will find out in later sections.

We note in passing that qualitatively, most of the considerations above carry over to high D cases with spherically symmetric impurity potentials, as (i) they do not mix different angular momentum sectors; and (ii) the asymptotic behavior of the energy eigenstates is uniquely determined by the corresponding phase shift. Quantitatively however, only the s-wave channel of the 3D case works exactly like the 1D case above, and in particular, in the low-energy limit where the $k$-dependence of the s-wave scattering phase shift takes the form
\begin{eqnarray}
\delta^{s}_k= k  \mathfrak{a}^{s},
\end{eqnarray}
where $ \mathfrak{a}^s$ is the s-wave scattering length, its effects are identical to those of a shift in the subsystem boundary. We thus consider in the following a special case where we have a 3D spherical that gives rise to s-wave scattering only, with the phase shifts taking the form above. In this case we find the s-wave energy eigenfunction in subsystem B takes the form $\sin(kr-ka^s)/r$, and the corresponding overlap matrix takes the form
\begin{eqnarray}
B^s_{kk'}&=&\int_L^{\infty}4\pi r^2[\sin(kr-k \mathfrak{a}^s)/r][\sin(k'r-k' \mathfrak{a}^a)/r]dr\nonumber\\
&=&\int_{L-a^s}^{\infty}\sin(kr)\cos(k'r)dr,
\end{eqnarray}
where we see the effect of $\mathfrak{a}^s$ is equivalent to shifting the boundary from $L$ to $L- \mathfrak{a}^s$. Also noticing the similarity to the 1D odd channel case, we conclude in this 3D case an s-wave scatterer induces a correction to entanglement entropy
\begin{eqnarray}
\delta EE(k_F, L)=-{ \mathfrak{a}^s\over 6L}+O(1/L^2).
\end{eqnarray}

\section{Numerical method}\label{method}

Coming back to the 1D case, we now
consider the Hamiltonian of a tight binding model with constant hopping amplitude $t$ and only one non-zero on-site energy. We refer to this as the single impurity (SI)
Hamiltonian:

\begin{equation}\label{Himp}
H_{SI}=-t\sum_{i=1}^{N} [c_{i}^{\dagger} c_{i+1} + c_{i+1}^{\dagger} c_i] + w c_{n}^{\dagger} c_{n},
\end{equation}
where we choose periodic boundary conditions (PBC) and there is one impurity with strength $w$ on a specific site $n$ (we specify $n$ below). $N$ is the total system size. The free fermion model
corresponds to the special case of $w=0$. We always choose $t=1$ and compare $w$ with $1$.

In this work we study the block entanglement entropy (EE) of  the single impurity Hamiltonian. To calculate EE, we consider a system with a finite size $N$ and with $N_F$ fermions. We divide the system into two parts, part $A$ from site $1$ to site $N_A$ and the rest as part $B$. In this paper we always choose $N$ to be even number and $N_A$ to be odd number such that we can put the impurity at the center of subsystem $A$, i.e. we choose index $n$ in Eq. (\ref{Himp}) to be the mid-point of the subsystem $A$.

For a system in a pure state $\ket{\psi}$, the density matrix is $\rho= \ket{\psi} \bra{\psi}$. The reduced density matrix of each subsystem (A or B) is obtained by
tracing over degrees of freedom of the other subsystem: $\rho^{A/B}=tr_{B/A} (\rho)$. The block EE between the two subsystems is
$EE=-tr(\rho^{A}\ln{\rho^{A}})=-tr(\rho^{B}\ln{\rho^{B}})$. For a single Slater-determinant ground state,
\begin{equation} \label{rho}
\rho^{A/B}=\frac{1}{Z} e^{-H^{A/B}}
\end{equation}
are characterized by the free-fermion {\em entanglement} Hamiltonians
\begin{equation}
H^{A/B} = \sum_{ij} h_{ij}^{A/B}  c_{i} ^{\dagger} c_{j},
\end{equation}
where $Z$ is determined by the normalization condition $tr \rho^{A/B}=1$. We calculate EE by using the method of Ref. \onlinecite{correl} by diagonalizing correlation
matrix of subsystem A
\begin{equation}
C_{mn}=\avg{c_{m} ^{\dagger} c_{n}},
\end{equation}
and find its eigenvalues $\zeta$'s. Then EE can be expressed as follows:
\begin{equation}
\text{EE}=-\sum_{l=1} ^{N_A} [\zeta_l \ln(\zeta_l)+(1-\zeta_l) \ln(1-\zeta_l)]\,.
\end{equation}

\section{Numerical Results and Discussion}\label{numeric}

The asymptotic behavior of EE in the one dimensional free fermion continuum system is as follows \cite{Ivanov}:
\begin{equation} \label{EEIvanov}
EE=\frac{1}{3} \ln{(2k_FL)} + \Upsilon + \orderof(1/L^2)
\end{equation}
where $L$ is the subsystem size and the Fermi wave vector $k_F=\pi N_F/N$ and the constant $\Upsilon\approx0.4950$. Our initial calculations focus on checking this asymptotic behavior for a finite and discrete system. In our numerical calculation, in order to
approach an infinite and continuous system, subsystem $A$ is chosen to be much smaller than the total system and $k_F$ is chosen to be small compared to $\pi$. More
specifically $N_A=N/10+1$ and $N_F=N_A$. Also we choose $N$ to be a multiple of $100$, in this way $N_A$ is an odd number and as explained before, we put the single impurity at
the middle of subsystem $A$. $N_F$ is also chosen to be odd to remove the ambiguity of which state the last fermion will occupy.

\begin{figure}
\includegraphics[width=0.5\textwidth]{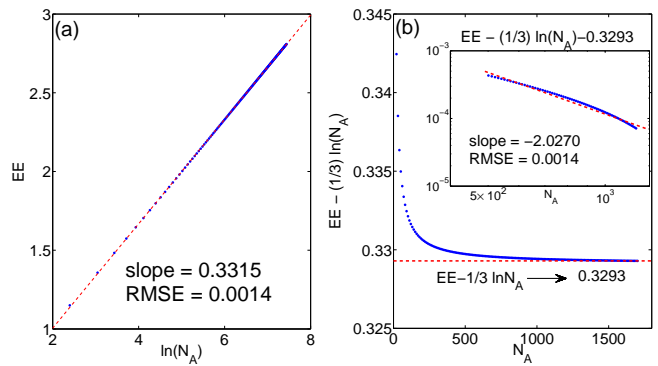}
\caption{\label{w0_3} [Color online] Panel (a): Free fermion entanglement entropy in log-linear scale. We set $N_A=N_F=N/10+1$, and $N$ goes from $100$ to $17000$ with
step of $100$. Blue dots are numerical results and the red line is fitted line with slope of $0.3315$ and root mean square error (RMSE)=$0.0014$. The slope is in
agreement with analytical value of $1/3$. Panel (b): main plot is $EE-1/3 \log{N_A}$ in linear -linear scale with same setting of panel (a). $EE-1/3 \log{N_A}$
approaches $0.3293$ for big subsystem size, close to the analytic value of $0.3401$. The inset plot is $EE-1/3\ln{N_A}-0.3293$ versus $N_A$ in a log-log scale. Slope of
the fitted line is $-2.0270$ and with RMSE=$0.0014$.}
\end{figure}

First, we calculate the EE of free fermion system (Eq. (\ref{Himp}) with $w=0$) and plot it versus $\ln{N_A}$ (Fig. \ref{w0_3}, panel (a)). The slope of the fitted line is
$0.3315 \approx 1/3$, confirming the logarithmic behavior in Eq. (\ref{EEIvanov}). Second, to obtain the constant term, we calculate $EE-1/3 \ln{N_A}$ and plot it versus
$N_A$ (Fig. \ref{w0_3}, panel (b)). $EE-1/3 \ln{N_A}$ approaches $0.3293$ for large subsystem size. This number is in agreement with the constant term of Eq.
(\ref{EEIvanov}) which is $1/3 \ln{(2\pi N_F/N)}+0.4950 \approx 0.3401$. And finally, to verify that there is subleading term on the order of $1/N_A^2$, we plot
$EE-1/3\ln{N_A}-0.3293$ versus $N_A$ in a log-log scale (see inset plot in Fig. \ref{w0_3}, panel (b)). Fitted line has slope of $\approx -2$ as expected. Note that
there is no oscillation in the behavior of EE in Fig. \ref{w0_3}.

\begin{figure}
\includegraphics[width=0.5\textwidth]{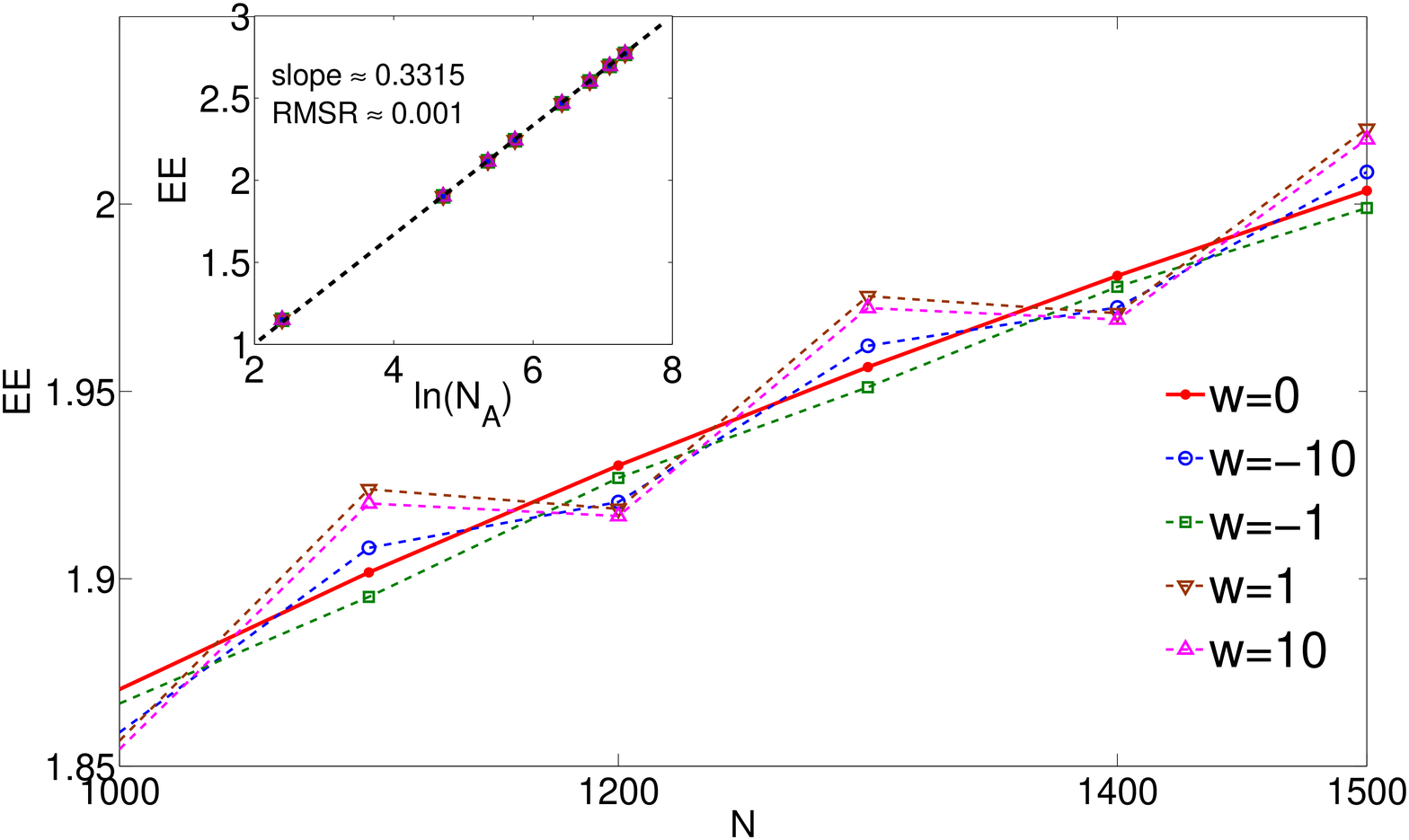}
\caption{\label{EE_allw2} [Color online] Main plot: entanglement entropy corresponding to different values of impurity strength  $w$. Free fermion case is plotted in
red. Small range of subsystem size is chosen to see the deviations of single impurity entanglement entropies. The inset is a log-linear plot of entanglement entropy for
different value of impurity strength  including $w=0$. Slope of fitted line is same for all value of $w$ and it is $0.3315$. In both plots, we set $N_F=N_A=N/10+1$. $N$
goes from $100$ to $17000$ with step of $100$. }
\end{figure}

Up to now we have verified Eq. (\ref{EEIvanov}), the EE of the free fermion model ($w=0$) numerically. Now, we focus on the effect of a single impurity on the EE. We
begin our calculations by comparing the free fermion and single impurity EE. In Fig. \ref{EE_allw2}, the EE of the free case, $EE_0$, and the EE
corresponding to different values of $w$'s, $EE_w$, are plotted. The inset of the plot shows the EE for different impurity strength  including $w=0$ in
log-linear scale, demonstrating that the single impurity does not change the log term of Eq. (\ref{EEIvanov}). This has been proven in a recent paper \cite{ossipov}.
However, when we zoom in (main plot) we see that the single impurity EE is oscillating around the free fermion EE. According to Fig. \ref{EE_allw2} it is evident that
the deviation from $EE_0$ is small compared to $EE_0$ at each point.

\begin{figure}
\includegraphics[width=0.5\textwidth]{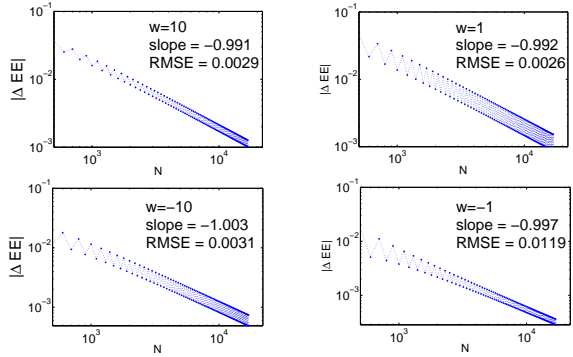}
\caption{\label{absdiff_loglog}The absolute value of difference between single impurity and free fermion entanglement entropy $|\Delta EE|$ for different values of $w$
in log-log scale. Beside oscillation, absolute difference goes to zero with slope approximately equals to $-1$. The slope and root mean square error (RMSE) are
indicated for each plot. We set $N_F=N_A=N/10+1$. $N$ goes from $100$ to $17000$ with step of $100$.}
\end{figure}

Now, to investigate the effect of the single impurity on the sub-leading terms, we calculate the difference between the single impurity EE and the free fermion EE:
$\Delta EE = EE_w-EE_0$. The absolute value of this difference, $|\Delta EE|$ for different values of $w$ is plotted in Fig. \ref{absdiff_loglog} in log-log scale (with
the same settings as in Fig. \ref{w0_3}). Numerical calculations of this figure show that although there is an oscillation, $|\Delta EE|$ goes to zero in log-log scale
with the slope of $\approx -1$. Thus, the single impurity adds a sub-leading term of order $1/N_A$ to Eq. (\ref{EEIvanov}) which has oscillating characteristics.

\begin{figure}
\includegraphics[width=0.5\textwidth]{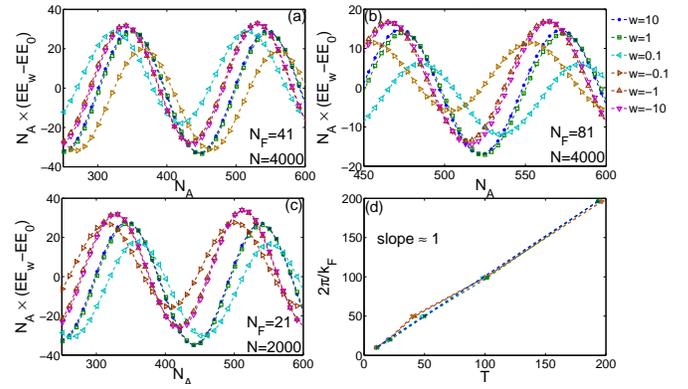}
\caption{\label{NAtimesDEE5} [Color online] Subsystem size $N_A$ times difference between single impurity and free fermion entanglement entropy, $N_A \times
(EE_w-EE_0)$ when $N_A$ is changing. We set $N=4000$, $N_F=41$ in panel(a) and $N=4000$, $N_F=81$ in panel (b) and $N=2000$, $N_F=21$ in panel (c). Panel (d):
$2\pi/k_F$ versus calculated oscillation periodicity $T$ for $N_A \times (EE_w-EE_0)$ when $N_A$ is changing. We set $N=5000, N_F=51,101,201,501,1001$. for different
values of impurity strength  $w$. Slope of the fitted line for different $w$'s is $\approx 1$. }
\end{figure}

Next, we study the quantity $N_A \times (EE_w-EE_0)$ to understand the oscillating characteristic of the sub-leading term of order $1/N_A$ better. We calculate this
quantity for two settings. In one setting, we set $N=4000, N_F=41,81$, and $N_A$ is changing, panel (a) and (b) Fig. \ref{NAtimesDEE5}. In the other one, we set $N=2000,
N_F=21$ and $N_A$ is changing, panel (c) in Fig. \ref{NAtimesDEE5}.

By examining panel (a) and (b) in Fig. \ref{NAtimesDEE5}, we see  $N_A \times (EE_w-EE_0)$ is oscillating with a constant periodicity.  The approximate periodicity ($T$) of
this oscillation is listed in Table  \ref{tablenanf} which shows that  $T$ does not depend on impurity strength  $w$ but it is inversely proportional to $k_F$.
Furthermore, the average value of the oscillation is shifted up and down for different values of $w$ (we denote this shift in Table \ref{tablenanf} by $\beta$) and the
amplitude of oscillation $\alpha$ is also $w$-dependent. By comparing the two cases of $N_F=41$ and $N_F=81$ we see that $\beta$ and $\alpha$ also depend on $k_F$. Thus, the $N_A
\times (EE_w-EE_0)$ term has an oscillating term plus a non-oscillating term. The amplitude and non-oscillating term depend on $w$ and $k_F$.

\begin{table}
\caption{\label{tablenanf} Numerical data observed in Fig. \ref{NAtimesDEE5}, panel (a) and (b): $N=4000$, $N_F=41,81$ and $N_A$ is changing. $\alpha$ is amplitude of
oscillation, $\beta$ is the shift in vertical direction, and $T$ is periodicity of term $N_A \times (EE_w-EE_0)$.}
\begin{tabular}{c|ccc||ccc}
$N_F=41$ &   &       &        & $N_F=81$  &    &  \\
\hline
$w$  &  $\alpha$  &  $\beta$ & $T$   &  $\alpha$  &  $\beta$ & $T$  \\
\hline
$10$ & $30.85$ & $1.98$ & $200$ & $16.05$ & $0.73$ & $100$ \\
$1$ & $30.38$ & $2.53$ & $200$ & $15.54$ & $1.21$ & $100$ \\
$0.1$ & $23.41$ & $5.96$ & $200$ & $8.87$ & $2.57$ & $100$ \\
$-0.1$ & $26.00$ & $-6.52$ & $200$ & $9.65$ & $-2.80$ & $100$ \\
$-1$ & $31.64$ & $-2.45$ & $200$ & $15.91$ & $-1.40$ & $100$ \\
$-10$ & $31.94$ & $-1.92$ & $200$ & $16.22$ & $-0.89$ & $100$ \\
\end{tabular}
\end{table}

To compare two cases with same $k_F$, we consider panel (a) and panel (c). In panel (a) we have: $N=4000, N_F=41, k_F \approx \pi/100$ and in panel (b) we have $N=2000,
N_F=21, k_F \approx \pi/100$. These two cases have equal $T$. We conclude that the oscillating term has the form of $\cos{(c k_F N_A+ \theta)}$, where $c$ is a
constant and $\theta$ is a phase shift. To measure $c$, we set $N=5000$ and for different $N_F=51,101,201,501,1001$, we calculate oscillation periodicity of $N_A \times
(EE_w-EE-0)$ term. Then we plot $2\pi/k_F$ versus $T$. As we can see in panel (d) of Fig. \ref{NAtimesDEE5}, the slope of the fitted line, $c$, is the same for different
$w$'s and is  $\approx1$. Thus the subleading term introduced by the single impurity has the following form:

\begin{equation}
\frac{\alpha\cos{( k_F N_A + \theta)} + \beta}{N_A}.
\end{equation}

In the following subsections, we discuss the behavior of the oscillating amplitude $\alpha$ and the non-oscillating term $\beta$ in different regimes of impurity strength.

\subsection{Large impurity strength}

First, we consider the case of $w \to \infty$. For a discrete lattice system, large on-site energy means that electrons are  banned from the impurity site. This corresponds to cutting the system at the impurity site (which is located at the middle of the subsystem). When we use a system with PBCs, cutting the system at the middle of the subsystem results in a system with open boundary conditions and half of the subsystem is located at the beginning and the other half of the subsystem is located at the end of the system. The calculation of the EE of a semi infinite lattice system has been done before \cite{fagotticalabrese}. In Ref. \onlinecite{fagotticalabrese} the subleading terms of the R\'enyi EE of order $n$ for a subsystem with length $\ell$
(where the subsystem starts at the beginning of system), $\Delta_n$, are calculated as follows: \begin{equation}
\Delta_n=\frac{2\sin{[k_F(2\ell +1)]}}{1-n}[2(2\ell+1)|\sin{k_F}|]^{-1/n} \frac{\Gamma{(1/2+1/2n)}}{\Gamma{(1/2-1/2n)}}.
\end{equation}

The von Neumann EE corresponds to $n \to 1$. By replacing $\ell \to (N_A-1)/2$ and considering two of these semi infinite systems, subleading term of EE of two semi
infinite subsystems for small $k_F$ is:

\begin{equation}\label{delta}
\Delta=\frac{\sin{(k_FN_A)}}{N_A k_F}
\end{equation}

Thus, in the very large impurity strength  limit (when $w \gg 1$), the subleading term goes to zero as $1/N_A$ and the oscillation amplitude $\alpha=1/k_F$ and
the non-oscillating term $\beta=0$.

\begin{figure}
\includegraphics[width=\linewidth]{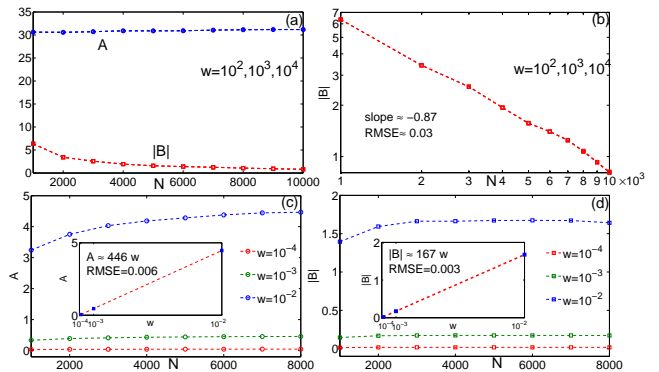}
\caption{[Color online] Panel (a): the oscillation amplitude $\alpha$ (blue) and non-oscillating term $\beta$ (red) of $N_A \times (EE_w-EE_0)$ for a constant $k_F=\pi/100$
corresponding to $w=10^2$, $10^3$, $10^4$. Panel (b): The non-oscillating term in a log-log scale. slope of the three $w$'s is approximately the same: $\approx -0.87$.
panel (c): the oscillation amplitude $\alpha$  of $N_A \times (EE_w-EE_0)$ for a constant $k_F=\pi/100$ corresponding to $w=10^{-2}$, $10^{-3}$, $10^{-4}$ from top to bottom. Inset plot is saturated value of $\alpha$ versus $w$. Panel(d): absolute value of the non-oscillating term $|\beta|$ of $N_A \times (EE_w-EE_0)$ for a constant $k_F=\pi/100$ corresponding to $w=10^{-2}$, $10^{-3}$, $10^{-4}$ from top to bottom. Inset plot is saturated value of $\beta$ versus $w$.}
\label{Absmalllargew}
\end{figure}

On the other hand, to compare our results with Eq. (\ref{delta}), we calculate the oscillation amplitude $\alpha$ and the non-oscillating term $\beta$ for large values of $w$ numerically. We calculate $\alpha$ and $\beta$  for a fixed $k_F=\pi/100$ and as we approach large system sizes. The results are plotted in Fig. \ref{Absmalllargew}, panel (a) for $w=10^2,10^3,10^4$. We see that  large $w$'s all lead to the same behavior. The oscillation amplitude is approximately constant $\approx 31.2$ which is close to the oscillation amplitude of Eq. (\ref{delta}): $1/k_F=1/(\pi/100)=31.8$. Moreover, Fig. \ref{Absmalllargew}, panel (b) demonstrates
the behavior of the non-oscillating term $|\beta|$ in a log-log scale. We see that the non-oscillation term goes to zero, which is in agreement with Eq. (\ref{delta}).

\subsection{Small impurity strength}
For small values of impurity strength, $w \ll 1$, numerical calculations (Fig. \ref{Absmalllargew}, panel (c) and panel (d)) show that the oscillation amplitude and the
non-oscillation term saturate when we approach large system size for a fixed $k_F=\pi/100$. This saturation occurs faster for smaller $w$'s. Saturated value of $\alpha$ and $|\beta|$ are plotted versus $w$ in inset plots. We see that both $\alpha$ and $|\beta|$ depend linearly on the impurity strength $w$ and as we expect $\alpha$ and $|\beta|$ go to
zero as $w \to 0$.

\subsection{Disorder on the order of $1$}
After discussing limiting case of very large and very small impurity strength, we now consider $w \sim \orderof(1)$. We claim that in this regime the analytic
calculation presented in section \ref{analytic} (which is for the continuum case) provides us with the mean value of the subleading term, i.e., the non-oscillating term. We use our lattice model to verify this by choosing a small $k_F$ to approach the continuum case. We will restrict the regime under consideration more precisely later. First, we calculate the phase shift in the case of a one dimensional free fermion lattice, where there is an impurity with strength of $w$ at the origin. The odd
sector of the Hilbert space does not see the impurity at the origin and thus there is no phase shift ($\delta_k^o=0$ and thus $\mathfrak{a}^o=0$), while for the even
sector a phase shift can be calculated as follows:

We write the discrete even wavefunction of a free fermion lattice in presence of a single impurity at the site with the index $j=0$ as:
\begin{equation}\label{psij}
    \psi_j^{even}=
\begin{cases}
    C_1 e^{ikja}+ C_2 e^{-ikja},    j<0 \\
    C_2 e^{ikja}+ C_1 e^{-ikja},    j>0
\end{cases}
\end{equation}

By writing the Schroedinger equation $H \ket{\psi}=\varepsilon \ket{\psi}$ and $\ket{\psi} =\sum_j \psi_j \ket{j}$, we can obtain the equation for the amplitudes:
\begin{equation}\label{tpsi}
-t(\psi_1+\psi_{-1})+w\psi_0=\varepsilon \psi_0
\end{equation}

By substituting the wavefunction from Eq. (\ref{psij}), we obtain:
\begin{equation}
C_2=-C_1 \frac{\frac{w-\varepsilon}{2t}-\cos{ka}+i\sin{ka}}{\frac{w-\varepsilon}{2t}-\cos{ka}-i\sin{ka}}.
\end{equation}
Now, if we write $C_2=e^{i \delta_k^e}C_1$, we have:
\begin{equation}
\tan{\delta_k^e}=\frac{2(\frac{w-\varepsilon}{2t}-\cos{ka})\sin{ka}}{(\frac{w-\varepsilon}{2t}-\cos{ka})^2-\sin^2{ka}},
\end{equation}
in the low energy limit where $k_F a \ll 1$ and thus $\varepsilon_F \approx -2t$, and also if we set $t=1, a=1$, we will have:

\begin{equation}
\tan{\delta_k^e}=\frac{4\frac{k}{w}}{1-4(\frac{k}{w})^2}.
\end{equation}

Writing the phase shift as $\delta_k^e=k\mathfrak{a}^e$ casts a condition on $w$: $k_F/w \ll 1$. In this regime the scattering length is:
\begin{equation}
\mathfrak{a}^e=\lim_{k \to 0}\frac{4k/w}{k}=\frac{4}{w}.
\end{equation}

On the other hand, writing Eq. (\ref{tpsi}) at the Fermi level, where $\varepsilon_F \approx -2t$, we have: $\psi_0 = \frac{t(\psi_1+\psi_{-1})}{w+2t}$. Continuity of
the $\psi$ yields the condition $w/t \ll 1$. Putting these two conditions on $w$, we have:
\begin{equation}\label{condkfw}
k_F \ll w \ll 1.
\end{equation}

Thus, by using the Eq. (\ref{eq:ContinuumEEChange}) in the regime of Eq. (\ref{condkfw}) the non-oscillating term is :
\begin{equation}\label{b}
|\beta|=\frac{\mathfrak{a}^e+\mathfrak{a}^o}{6}=\frac{2}{3w}.
\end{equation}

\begin{figure}
\includegraphics[width=\linewidth]{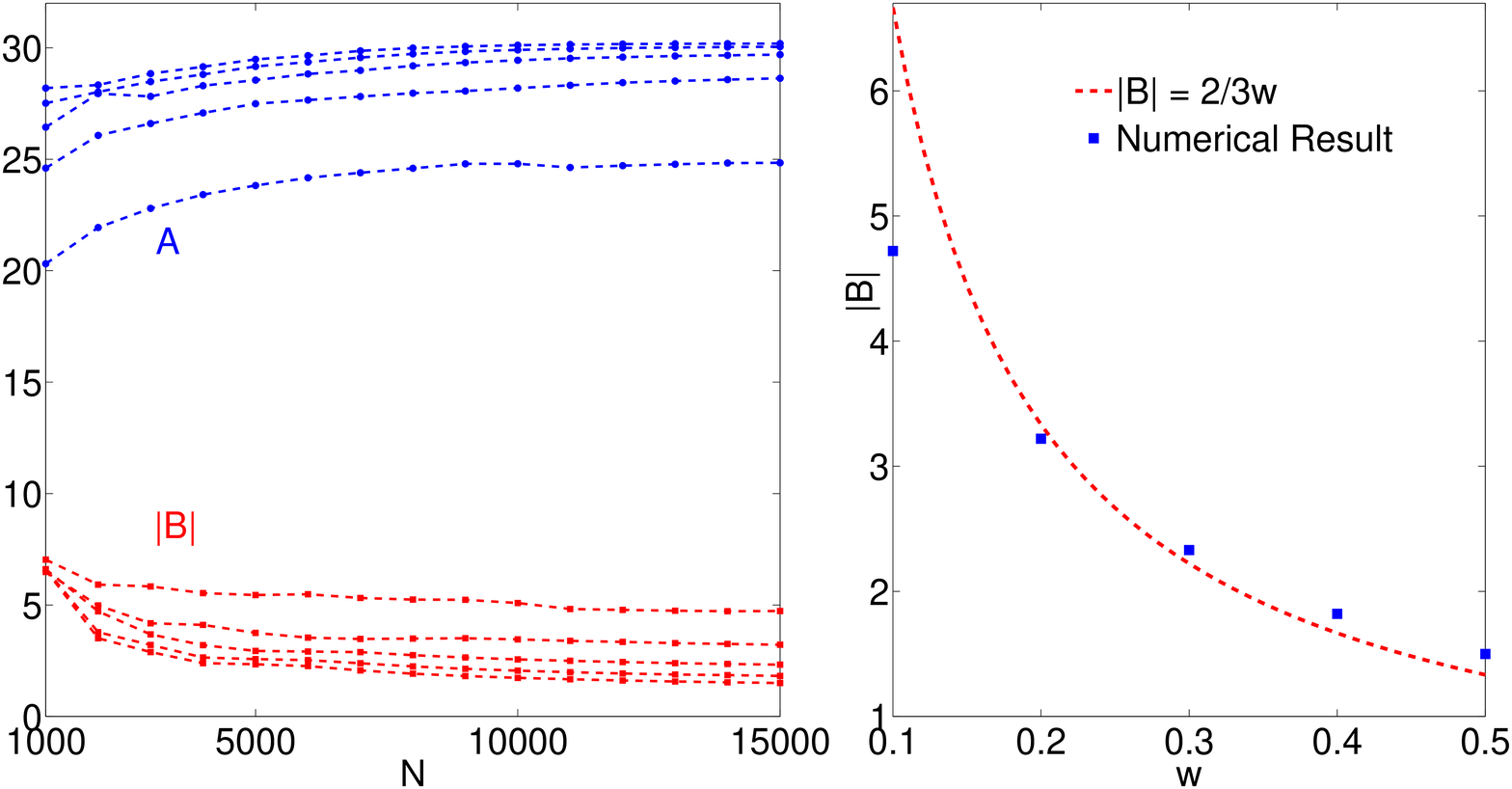}
\caption{[Color online] Panel (a): the oscillation amplitude $\alpha$ (blue) and non-oscillating term $\beta$ (red) of $N_A \times (EE_w-EE_0)$ for a constant $k_F=\pi/100$
corresponding to $w=0.1$, $0.2$, $0.3$, $0.4$, $0.5$, from bottom to top for $\alpha$ and from top to bottom for $\beta$. Panel (b): Comparison between saturated valued of $\beta$ and Eq. (\ref{b}). }
\label{ABorder1w}
\end{figure}

Now, we compare Eq. (\ref{b}) with our numerical calculation of non-oscillating term. To approach the continuum case we choose a small $k_F= \pi/100$ and to fulfill the requirement in Eq. (\ref{condkfw}), we choose $w=0.1,0.2,0.3,0.4,0.5$. The results for the oscillating amplitude $\alpha$ and the non-oscillating term $\beta$ are plotted in Fig. \ref{ABorder1w}, panel (a). The saturated value for  $\beta$ and its behavior according to Eq. (\ref{b}) are plotted in Fig. \ref{ABorder1w}, panel (b) for comparison. As we can see in this plot, numerical results are close to the value calculated by Eq. (\ref{b}). Moreover, the saturated value of the oscillation amplitude becomes closer to the value $1/k_F=31.8$ for larger $w$'s.

\section{Concluding Remarks}
\label{con}

In this work we have studied in detail the effects of a single impurity on the entanglement entropy of a 1D free Fermi gas, where we find subleading corrections that scale inversely with the subsystem size, using a combination of analytical and numerical techniques. This subleading term has an oscillating and a non-oscillating part. The period of oscillation is independent of impurity strength, but the amplitude of the oscillation $\alpha$, and the non-oscillation term $\beta$, both depend on impurity strength $w$: in the limit of large $w$, the amplitude approaches the value $1/k_F$ and the non-oscillating term goes to zero. On the other hand, in the limit of small $w$, both $\alpha$ and $\beta$ are proportional to $w$. We also found that when $w$ is in the  regime of Eq. (\ref{condkfw}), our analytical calculations predict the average $\beta$ of the subleading term, Eq (\ref{b}).

Our results may be surprising from the perspective of scaling analysis. An exactly marginal operator is characterized by a dimensionless coupling constant, which, in this case, is the phase shift at the Fermi energy. One would expect this operator to induce a correction to the entanglement entropy that is {\em independent} of the subsystem size $L_A$. Our results suggest that the correction due to the phase shift is {\em identically zero}! This has been confirmed by an explicit CFT calculation\cite{AffleckPrivate}. The corrections we have found, therefore, come from the derivative of the phase shift with respect to the Fermi wave vector, which is actually an {\em irrelevant} operator with scaling dimension one. This explains the $1/L_A$ dependence of these corrections\cite{AffleckPrivate}.

Our results are also relevant for the free Fermi gas in higher dimensions. This is particularly true for spherically symmetric impurity potentials and a partition that respects rotational symmetry. In this case different angular momentum channels decouple, and each channel behaves as a half-infinite 1D chain. The correction to the EE due to the impurity is thus the sum of its correction to each (1D) channel.

Simple as they may be, free fermion systems are in fact highly non-trivial from an entanglement point of view. They represent the first known examples of area law violation in ground state entanglement entropy,\cite{GioevKlich,wolf,spitzer} which were found in other systems only recently.\cite{dingprx12,LaiYangBonesteel} Very recent work has also elucidated entanglement properties of highly excited states in such systems.\cite{LaiYang} Extensive work on entanglement in disordered free fermion systems has also been carried out.\cite{Peschel, Hughes, PourYang, Potter, PourZhangYang,igloi, eislerpeschel1,eislerpeschel2,berkovits} The present work is a new addition to this body of results.

We note that a single impurity in a free 1D Fermi gas is very special. Interactions change its scaling dimension\cite{KaneFisher92}, making it a relevant perturbation for repulsive interaction, and an irrelevant one for attractive interaction. We thus expect finite corrections to the entanglement entropy for the former, and a subleading one for the latter, with an exponent varying continuously with interaction strength. While these conclusions follow straightforwardly from scaling analysis\cite{affleckreview}, it is not immediately clear if there will be oscillatory contributions, and if so, with what period. These questions deserve further investigation.

\acknowledgments
We thank Ian Affleck and Nick Read for illuminating discussions.
This work was supported by NSF Grants DMR-1442366 (MP and KY) and No. DMR-1206781 (AS).

\end{document}